\documentclass[prl,twocolumn,superscriptaddress,showpacs]{revtex4-1}
\usepackage{verbatim}
\usepackage{graphicx}
\usepackage{color}
\usepackage{multirow}
\usepackage{subfig}
\usepackage{float}
\usepackage{caption}
\usepackage{amsxtra}
\usepackage{amscd}
\usepackage{amsthm}
\usepackage{amsfonts}
\usepackage{comment}
\usepackage{slashbox}		
\usepackage[dvipsnames]{xcolor}
\usepackage{tikz}	

\newcommand{\note}[1]{$\color{red}{\bullet}$}

\newcommand{\bComment}{}

\DeclareCaptionLabelFormat{andtable}{#1~#2  \&  \tablename~\thetable}

\begin{document}

\title{Entanglement Spectrum of Composite Fermion States in Real Space} 

\author{Iv\'{a}n D. Rodr\'{i}guez}
\affiliation{SISSA, Via Bonomea 265, 34100, Trieste, Italy}
\affiliation{INFN, Sezione di Trieste, Italy}
\author{Simon C. Davenport}
\author{Steven H. Simon}
\affiliation{Rudolf Peierls Centre for Theoretical Physics, 1 Keble Road, Oxford, OX1 3NP, UK}
\author{J.~K. Slingerland}
\affiliation{Department of Mathematical Physics, National University of Ireland, Maynooth, Ireland}
\affiliation{Dublin Institute for Advanced Studies, School of Theoretical Physics, 10 Burlington Rd, Dublin, Ireland}
 
\date{\today}

\begin{abstract}
We study the entanglement spectra of many particle systems in states which are closely related to products of Slater determinants or products of permanents, or combinations of the two. Such states notably include the Laughlin and Jain composite fermion states which describe most of the observed conductance plateaus of the fractional quantum Hall effect.  We identify a set of  'Entanglement Wave Functions' (EWF), for subsets of the particles, which completely describe the entanglement spectra of such product wave functions, both in real space and in particle space. A subset of the EWF for the Laughlin and Jain states can be recognized as Composite Fermion states. These states provide an exact description of the low angular momentum sectors of the real space entanglement spectrum (RSES) of these trial wave functions and
a physical explanation of the branches of excitations observed in the RSES of the Jain states.



\end{abstract}

\maketitle

In studying quantum states of many particle systems, it is often of interest to split the system into subsystems and focus on the reduced density matrix $\rho_A$ of one of these subsystems. Various \emph{entanglement spectra}\cite{Li08} collect the eigenvalues and eigenstates of such reduced density matrices for different \lq cuts' of the initial system into an $A$-subsystem and a $B$-subsystem. Two very natural cuts for an $N$ particle system are the \emph{particle cut}, for which the $A$-system simply consists of particles $1$ through $N_A$ (The $B$-system has the remaining particles) and the \emph{real space cut}, where the $A$-system consists of particles within a certain spatial domain and the $B$-system has the particles in the complementary domain. These cuts lead to the particle entanglement spectrum (PES) and the real space entanglement spectrum (RSES).  If the full system is in a state $\psi$, then we may write $\rho_{A}(Z_{A},Z^{\prime}_{A})=\int_{D_{B}}\psi^{*}(Z_A,Z_{B})\psi(Z^{\prime}_{A},Z_{B})\, dZ_{B}$, where $Z_{A}$, $Z_{A'}$ and $Z_{B}$ are sets of coordinates for particles in the $A$ and $B$ subsystems. For the particle cut, the domain $D_{B}$ is all of space and $Z_{A}$ and $Z_{A'}$ contain arbitrary positions. For the real space cut, $D_{B}$ is the chosen domain for the particles in the $B$-system, while the positions $Z_{A}$ and $Z_{A'}$can be restricted to lie in $D_{A}$, the complement of $D_{B}$. We should expected the PES and RSES to be adiabatically connected, as we can continuously grow the domains $D_{A}$ and $D_{B}$ for the $A$ and $B$ systems, allowing them to overlap and eventually to encompass all space. In the context of fractional quantum Hall (FQH) systems, the particle cut was studied in Refs.~\onlinecite{Haque07,Zozulya07,Sterdyniak11} and the real space cut 
in Refs.~\onlinecite{Sterdyniak12,Dubail12,Rodriguez12} and for filling $\nu=1$ in  Refs.~\onlinecite{Rodriguez09,Rodriguez10}. The eigenspaces of $\rho_A$ at fixed angular momentum were indeed found the be of equal dimension for PES and RSES, though the eigenvalues are different.

The most ubiquitous states appearing in many particle physics are Slater determinants (for fermions) and permanents (for bosons). These have particularly simple reduced density matrices due to their factorization properties. For a Slater determinant $S_{\phi_1...\phi_N}(z_1,...,z_N)$, involving $N$ particles and $N$ orbitals $\phi_1...\phi_N$, we have 
\begin{equation}
\label{eq:slaterfac}
\begin{array}{l}
S_{\phi_1...\phi_N}(z_1...z_N)=
\frac{1}{N_{A}!N_{B}!}\sum_{\sigma\in S_{N}}\epsilon(\sigma)\epsilon(\sigma|_{A})\epsilon(\sigma|_{B}) \\
\times S_{\phi_{\sigma(1)}...\phi_{\sigma(N_{A})}}(z_1...z_{N_A})S_{\phi_{\sigma(N_{A}+1)}...\phi_{\sigma(N)}}(z_{N_A+1}...z_N)
\end{array}
\end{equation}
Here the $\epsilon$-factors come from the usual antisymmetric tensors, that is, $\epsilon(\sigma)$ is the sign of the permutation $\sigma\in S_N$ and $\epsilon(\sigma|_{A})$ and $\epsilon(\sigma|_{B})$ give the sign of this permutation when it is restricted to $\{1...N_A\}$ and $\{N_{A}+1...N\}$. Note that the summands depend only on the sets $\{\sigma(1)...\sigma(N_A)\}$ and the factor $\frac{1}{N_{A}!N_{B}!}$ can be removed if we change the sum to a sum over all the possible choices of $N_A$ out of $N$ orbitals. We see then that the Slater determinant splits into a sum of products of Slater determinants for the subsystems,  with complementary occupation of the orbitals in the factors. For permanents one gets a similar result without the $\epsilon$-factors. It is now easy to write the density matrix explicitly in terms of the determinants (or permanents) for the $A$-system and one finds that the eigenvalues of $\rho_A$ are all equal and the eigenfunctions in the PES are precisely all the determinants (or permanents) that occur for the $A$-system. The RSES will usually be less trivial: in particular, the eigenstates of $\rho_A$ which have the $N_A$ particles of system $A$ concentrated in the domain $D_{A}$ will typically have the largest eigenvalues (or the lowest entanglement energies). 

The low energy states of strongly correlated systems are typically superpositions of many Slater determinants or permanents and have much more complicated and interesting entanglement spectra. Nevertheless, many trial wave functions for such systems, notably the Laughlin and Jain wave functions for FQH states, are products of Slater determinants, or projections of such products. Wave functions of this type have factorizations similar to Eq.~(\ref{eq:slaterfac}) and their PES and RSES can be completely described by a set of \emph{entanglement wavefunctions} (EWF), which are themselves products of Slater determinants.  To show this explicitly, we consider a general product of Slater determinants $\psi(z_1...z_N)=\prod_{i=1}^{m} S^{(i)}(z_1...z_N)$. Here, each of the $S^{(i)}$ is a slater determinant involving a set of $N$ orbitals $\phi^{(i)}_{1},\ldots,\phi^{(i)}_{N}$. It is clear from the expansion in Eq.~(\ref{eq:slaterfac}) that we can write 
\[
\psi(Z_A,Z_B)=\sum_{k} s_k \prod_{i=1}^{m}S^{(k^{A}_{i})}(Z_A)\prod_{i=1}^{m} S^{(k^{B}_{i})}(Z_B).
\]
Here the summation variable $k$ runs through all splittings of the sets of orbitals $\{\phi^{(i)}_{1},\ldots,\phi^{(i)}_{N}\}$ into subsets of $N_A$ and $N_B$ orbitals associated with the $A$ and $B$ subsystem. We may write $k=(k^{A}_{1},k^{B}_{1},...,k^{A}_{m},k^{B}_{m})$, where $k^{A}_{i}$ is the subset  of $\{\phi^{(i)}_{1},\ldots,\phi^{(i)}_{N}\}$ consisting of the $N_A$ orbitals which are assigned to the $A$-system and $k^{B}_{i}$ contains the remaining $N_B$ orbitals in $\{\phi^{(i)}_{1}\ldots,\phi^{(i)}_{N}\}$. The constants $s_{k}$ are the combined signatures of permutations that appear and hence $s_{k}\in\{1,-1\}$. The entanglement wave functions are the products  $\xi_{k}(Z_A):=\prod_{i=1}^{m}S^{(k^{A}_{i})}(Z_A)$ and $\zeta_{k}(Z_B):=\prod_{i=1}^{m} S^{(k^{B}_{i})}(Z_B)$ appearing in this factorization. 
We may now write
\begin{eqnarray}
\label{eq:rhoA}
&&\rho_A(Z_A,Z'_A)=\sum_{k,l}Q_{k,l\,}\xi_{k}(Z_{A})\xi^{*}_{l}(Z'_{A}), ~~~\mathrm{with}\nonumber \\
&&Q_{kl}=s_{k}s_{l} \int_{D_{B}} \zeta_{k}(Z_{B})\zeta^{*}_{l}(Z_{B}) \,dZ_{B}.
\end{eqnarray}
If we write the eigenvalues and eigenvectors of $\rho_{A}$ as $\lambda_{p}$ and $g_{p}$, so $\int \rho_{A}(Z_{A},Z'_{A})g_{p}(Z'_{A})\,dZ'_{A}=\lambda_{p}g_{p}(Z_{A})$, then we see that, for any $g_p$ with $\lambda_{p}\neq 0$, we have 
\begin{equation}
\label{eq:EWFexpansion1}
g_{p}(Z_A)=\sum_{k} \eta_{pk}\xi_{k}(Z_{A}),
\end{equation}
with
\begin{equation}
\label{eq:EWFexpansion2}
\eta_{pk}=\frac{1}{\lambda_{p}}\sum_{l}Q_{kl}\int_{D_{A}} \xi^{*}_{l}(Z'_{A}) g_{p}(Z'_{A})\,dZ'_{A}.
\end{equation}
%
%
Hence, we see that the states $g_p$ in the entanglement spectrum can be written as linear combinations of the EWF. 
Substituting the expression for $g_p(Z_A)$ from Eq.~(\ref{eq:EWFexpansion1})  into the expression for $\eta_{pk}$,  we obtain
\begin{equation}
\label{eq:EWFeigenvalues1}
\sum_{l} M_{il} \eta_{k l} = \lambda_k \eta_{k i}, 
\end{equation}
with
\begin{equation}
\label{eq:EWFeigenvalues2}
M_{il}=\sum_j Q_{ij} \int_{D_A}  \xi^*_j(Z_A') \xi_l(Z_A')\, d Z'_A.
\end{equation}
%
Thus, the nonzero eigenvalues $\lambda_p$ of $\rho_{A}$ are also eigenvalues of a finite dimensional matrix, $M$, given in terms of the overlap matrix of the EWF. In fact, there is a one to one correspondence between eigenvectors $\eta_k$ of $M$ with nonzero eigenvalues and corresponding eigenvectors $\sum_i{\eta_{ki}\xi_{i}}$ of $\rho_A$ (in particular, the rank of $M$ is equal to the rank of $\rho_A$). Hence the entanglement spectrum can be described, completely, in terms of the EWF: the nonzero eigenvalues of $\rho_A$ are given by (\ref{eq:EWFeigenvalues1}) and the corresponding eigenvectors by (\ref{eq:EWFexpansion1}), where the $\eta_{p}$ are eigenvectors of $M$.

All calculations above work equally well no matter what the domain of integration $D_{B}$ is, so the EWF play the same role both for the PES and for the RSES with any choice of domains for the $A$ and $B$ systems. One may also use products of permanents instead of Slater determinants or even mix permanents and determinants. Often, the state $\psi$ and the orbitals $\phi^{(i)}_{j}$  are eigenstates of some symmetry, e.g.~momentum $p$ or angular momentum $L$, so that $p_{\psi}=\sum_{i,j}p_{\phi^{(i)}_{j}}$ or $L_{\psi}=\sum_{i,j}L_{\phi^{(i)}_{j}}$. The corresponding EWF are then also eigenfunctions of the symmetry and if the choice of domain $D_{B}$ respects the symmetry, $\rho_A$ is block diagonal with blocks labeled by eigenvalues of the symmetry  and the calculation above can be performed within every block, using only the EWF with the appropriate eigenvalue of the symmetry. 

Usually the number of EWF $\xi_i$ is larger than the number of eigenstates $g_p$ of $\rho_A$, but the $\xi_i$ are not linearly independent. We will now argue that we should expect that the full space spanned by the $\xi_i$ is needed to describe the ES. From Eq.~(\ref{eq:EWFeigenvalues2}), we see that  any linear relation between the EWF gives rise to a zero eigenvector of $M$: if  $\sum_{i} \eta_i \xi_i=0$ then $M \eta=0$. Hence the rank of $M$ is less than or equal to the dimension of the space spanned by the EWF.  If the rank of $M$ is in fact smaller, there must be a vector $\eta$ such that $M\eta=0$ but $\Xi_{\eta}=\sum_{i} \eta_i \xi_i\neq 0$. In this case, $\Xi_{\eta}$ is orthogonal to $\sum_{j} Q^{*}_{ij} \xi_j$ for all $i$.
We now show that $\Xi_{\eta}$ with these properties do not exist if the $\xi_i$ are holomorphic functions of the $z_{i}$. In Ref.~\onlinecite{Rodriguez12}, three of us showed that, for holomorphic states, the rank of $\rho_A$ does not depend on the domain $D_{A}$. 
Actually the states considered were not quite holomorphic --- the property we used is that they are determined by their values on any open subset of their domain. This allows for example FQH trial wave functions built from orbitals in the lowest Landau level, which are holomorphic up to nonzero geometrical factors (gaussians for a system on a disk). 
We will use a straightforward generalization of this result. Instead of having a single domain $D_{A}$ for all particles of the $A$-system, we can allow individual (open) domains  $D_{A,i}$ in space for each of these particles. The rank of $\rho_A$ and hence $M$ still does not depend on the choice of these domains. Consider domains $D_{A,i}$ which are very small neighborhoods of points of space.
We take $D_{B}$ to be the full space. We now get, for $p=(z^{A}_{1},z^{A}_{2},...,z^{A}_{N_{A}})$,
\[
\sum_{j} \! Q_{ij} \!\! \int_{D_{A}} \!\!\!\!\! \xi^{*}_j (Z_A) \, \Xi_{\eta}(Z_A)\,dZ_{A}\approx \mathcal{A}\sum_{j} Q_{ij} \xi^{*}_{j}(p) \sum_{l}\eta_{l} \xi_l (p) 
\]
Here, $\mathcal{A}$ is the product of the areas of the $D_{A,i}$. The expression above should be zero by the property that $\Xi_{\eta}$  is orthogonal to $\sum_{j} Q^{*}_{ij} \xi_j$. However, $\sum_{j} Q_{ij} \xi^{*}_j$ is antiholomorphic and clearly nonzero for almost all $p$ (or else $M$ would be zero). Hence we find that $\sum_{l}\eta_{l} \xi_l (p) =0$ for almost all $p$ and hence $\sum_{l}\eta_{l} \xi_l=0$, since this expression is holomorphic. Indeed, no $\Xi_{\eta}\neq 0$ exists. The rank of $M$ thus equals the number of independent EWF and we conclude that, when the EWF are holomorphic, the entanglement spectrum spans exactly the same vector space as the EWF.

We now apply the above results to the Laughlin and Jain FQH states. These can be written in the form 
\begin{equation}
\label{eq:LJ}
\psi_{LJ}(z_1...z_N)=\mathbf{P}\left[\chi_{n}(z_1...z_N)(\chi_{1}(z_1...z_N))^{p}\right].
\end{equation}
Here, the $z_i$ are complex coordinates on a two dimensional space (usually the plane, sphere or torus) and $\chi_{n}$ is the Slater determinant describing $n$ fully filled Landau levels of orbitals on this space. Note that the number of orbitals in the Landau levels is adjusted to the number of particles. The single Landau level of the $\chi_1$ factors has $N$ orbitals,  while the $n$ Landau levels in the  $\chi_n$ factor have on average $N/n$ orbitals. One may think of these as Landau levels for composite fermions\cite{JainBook}, which experience a reduced magnetic flux, reducing the number of orbitals available.   
The operator $\mathbf{P}$ is the orthogonal projection onto the lowest Landau level for the full system. Eq.~(\ref{eq:LJ}) gives an excellent description of the ground state of FQH systems at filling $\nu=\frac{n}{pn+1}$. Excited states above this ground state can be described by modifying the first Slater determinant ($\chi_n$) so that it leaves some of the orbitals in the $n$ Landau levels unoccupied and/or includes some occupied orbitals in the $(n+1)^{\rm th}$ Landau level. The factor $(\chi_{1})^{p}$ is often written in the form $\prod_{i<j}(z_i-z_j)^{p}$. It is not involved in excitations and can be thought of as a device to attach flux to the composite fermions. We call all states of the form just described, with or without excitations, composite fermion (CF) states. The Laughlin states are the case $n=1$.

CF states are usually not products of Slater determinants, unless $\mathbf{P}$ acts trivially, which happens for the Laughlin ground states and some of their excitations. However, since the orthogonal projection $\mathbf{P}$ factors into single particle projections, we can still write 
\[
\psi_{CF}=\sum_{k} s_k 
\mathbf{P}\left(\prod_{i=1}^{p+1} S^{(k^{A}_{i})}(Z_A)\right)
\mathbf{P}\left(\prod_{i=1}^{p+1} S^{(k^{B}_{i})}(Z_B)\right),
\]
where the $p+1$ Slater determinants are now built from subsets of the orbitals involved in $\chi_{n}$ and in the $p$ factors of $\chi_1$.  We see then that it makes sense to define the EWF for the  wave functions in Eq.~(\ref{eq:LJ})  to be projected products of Slater determinants,  $\xi_{k}(Z_A):=\mathbf{P}\left(\prod_{i=1}^{p+1}S^{(k^{A}_{i})}(Z_A)\right)$ and $\zeta_{k}(Z_B):=\mathbf{P}\left(\prod_{i=1}^{p+1} S^{(k^{B}_{i})}(Z_B)\right)$. With this definition,  we can still apply the results (\ref{eq:rhoA})--(\ref{eq:EWFeigenvalues2}). In fact, FQH systems have symmetries which allow for a refined version of the construction above. E.g.~for the system on a sphere, the angular momentum component $L_{z}$ is conserved, even when the system is split into complementary $A$ and $B$ domains, as long as the domain boundary is a horizontal circle (e.g.~the equator of the sphere). Since $\mathbf{P}$ commutes with $L_z$,  $\psi_{CF}$  and the EWF are eigenfunctions of $L_z$ and  $\rho_A$ is block diagonal with blocks $\rho_{A}^{L^{A}_{z}}$ labeled by the total angular momentum $L^{A}_{z}$ of the particles in the $A$-system. We then get the results (\ref{eq:rhoA}) through (\ref{eq:EWFeigenvalues2})  for each block.

We note now that a subset of the EWF for the Laughlin and Jain states are themselves CF states. We will call these states the CF-EWF. The CF-EWF are those EWF for which the $p$ factors which come from the $(\chi_1)^{p}$ factor of $\psi_{CF}$ all have precisely the $N_A$ orbitals with the lowest angular momenta filled. This means that the product of these factors can be written as $\prod_{i<j=1}^{N_{A}}(z_i-z_j)^{p}$ and the factors are thus the usual flux attachment factor for the particles of the $A$-system. Clearly, in the expansion of $\psi_{CF}$, terms in which the $A$-system is in a CF-EWF have the $B$-system in a similar state where the $N_{B}$ highest angular momenta are filled in the Slater determinants that come from the $p$ factors of $\chi_1$. 
The CF-EWF are of particular interest when studying the low angular momentum sectors of the ES. For the RSES, low angular momentum corresponds with low entanglement energy, since states with low angular momentum $L^{A}_{z}$ have the highest probability density to find the particles $1...N_A$ in the domain $D_{A}$ (i.e. near the north pole of the sphere) and hence also the highest probability density for finding particles $N_{A}+1...N$ in $D_{B}$.
It is natural to guess that the CF-EWF give a good description of the low-energy part of the RSES of the Jain states from the point of view that the RSES should be similar to the edge spectrum of the parent state and this was our motivation for introducing these states very briefly in Ref.~\onlinecite{Rodriguez12}, without reference to the full set of EWF.
We may in fact expect that, at low angular momenta, the CF-EWF span the same space as the full set of EWF. This intuition comes from considering localized excitations of CF-states. These are coherent states built from the excitations we have described. A local excitation which exists for any FQH state is a Laughlin quasihole at position $w$, which can be introduced by multiplying the wave function by a factor $\prod_{j}(z_j-w)$. This excitation can be viewed as a superposition of Slater determinants with non-minimal angular momentum either at $\nu=n$ or at $\nu=1$. However, in the case of $\nu=1$ we can invert the construction; if we introduce a higher angular momentum orbital in one of the $\chi_1$ related factors of the EWF, this can always be thought of as a superposition of Laughlin quasiparticles. As a result, we could have created the higher angular momentum state also by introducing higher angular momentum orbitals in the $\nu=n$ factor of the EWF. It thus seems naively that the CF-EWF should fully describe the ES of the system. However, it is clear that the correspondence just described must break down beyond some maximal $L^{A}_{z}$; once the $N_A$ orbitals in the $\chi_n$ factor of the EWF are precisely the largest angular momentum orbitals in the $n$ CF LLs, the only way to raise the angular momentum further is to change the orbitals in the $\chi_1$ factors. Still, we expect that the set of angular momenta where the spaces of EWF and CF-EWF coincide increases with $N$ (with $N_A/N$ fixed while taking this limit), so that eventually, for any given excess in $L_{z}^{A}$ above the minimal $L_{z}^{A}$, the RSES in the limit $N\rightarrow \infty$ (with $N_{A}/N$ nonzero) may be obtained from the CF-EWF. 


We now present a comparison between the CF-EWF construction and the RSES of CF states on the sphere calculated using the method introduced in Ref.~\onlinecite{Rodriguez12}.
In Table \ref{tab:12and23count} we show the numbers EWF and CF-EWF (both the total numbers and the numbers of independent states), for different $L_z^A$ sectors,  for the $\nu=\frac{1}{2}$ Laughlin and $\nu=\frac{2}{3}$ Jain states on the sphere. We also give the number of independent states in the entanglement spectra for these states. 
As expected, the numbers of independent EWF match perfectly with the counting of the ES and the number of independent CF-EWF matches the EWF and ES at low angular momenta. We note that at low angular momenta, the CF-EWF are all linearly independent. In fact for the $\nu=\frac{1}{2}$ state, they are independent at any $L_{z}^{A}$ and it is not difficult to see that this is true for the CF-EWF of any state where the projection $\mathbf{P}$ acts trivially. More generally, we conjecture that the CF-EWF at any  given excess in $L_{z}^{A}$ above the minimal $L_{z}^{A}$ become linearly independent when $N\rightarrow\infty$ (keeping $0<N_A/N<1$ in the limit). 

\begin{table}
\begin{tabular}{c}
\resizebox{8.9cm}{!}{
\begin{tabular}{|c|*{11}{c}c|}
\hline 
   $L_z^A$ & -64 & -63 & -62 & -61 & -60 & -59 & -58 & -57 & -56 & -55 & -54 & -53  \\  \hline
   ES & 1 & 1 & 2 & 3 & 5 & 7 & 11 & 15 & 22 & 29 & 40 & 52  \\ 
   EWF    & 1 & 1 & 2(3) & 3(5) & 5(11) & 7(18) & 11(34) & 15(55) & 22(95) & 29(148) & 40(238) & 52(360)   \\
   CF-EWF & 1 & 1 & 2 & 3 & 5 & 7 & 11 & 15 & 22 & 28 & 38 & 48  \\ 
\hline 
\end{tabular}
}
\\
~\vspace*{-2mm}
\\
\resizebox{8.9cm}{!}{
\begin{tabular}{|c|*{7}{c}c|}
\hline 
   $L_z^A$ & -61 & -60 & -59 & -58 & -57 & -56 & -55 & -54   \\ \hline   
   ES & 1 & 4 & 9 & 20 & 40 & 72 & 121 & 194 \\
   EWF & 1  & 4 (5) & 9 (15) & 20 (40) & 40 (97) & 72 (212) & 121 (435) & 194 (843) \\
   CF-EWF  & 1 & 4 & 9 & 20 & 40(42) & 70(76) & 115(131) & 176(212)  \\   
\hline 
\end{tabular}
}
\end{tabular}
\caption{{\small ES, EWF and CF-EW countings. Numbers of independent states given. Total numbers of states in brackets, if different.
\emph{Top:} $\nu=1/2$ Laughlin state for $N=16$. \emph{Bottom:} $\nu=2/3$ Jain state for $N=18$.  }}
\label{tab:12and23count}
\end{table}

\begin{figure}[htb]
\begin{center}
\includegraphics[width= 4cm, height=6cm]{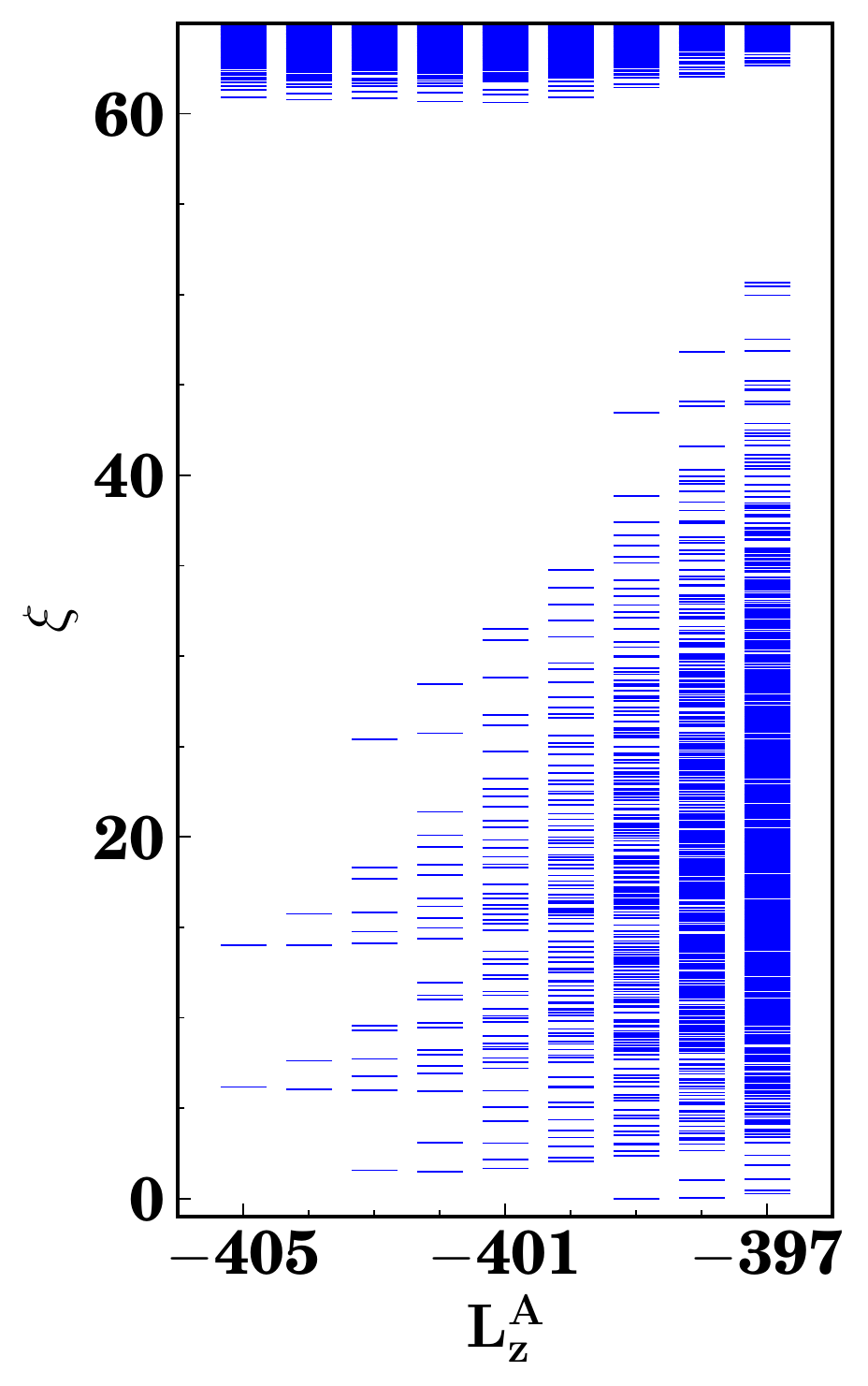} \includegraphics[width= 4cm, height=6cm]{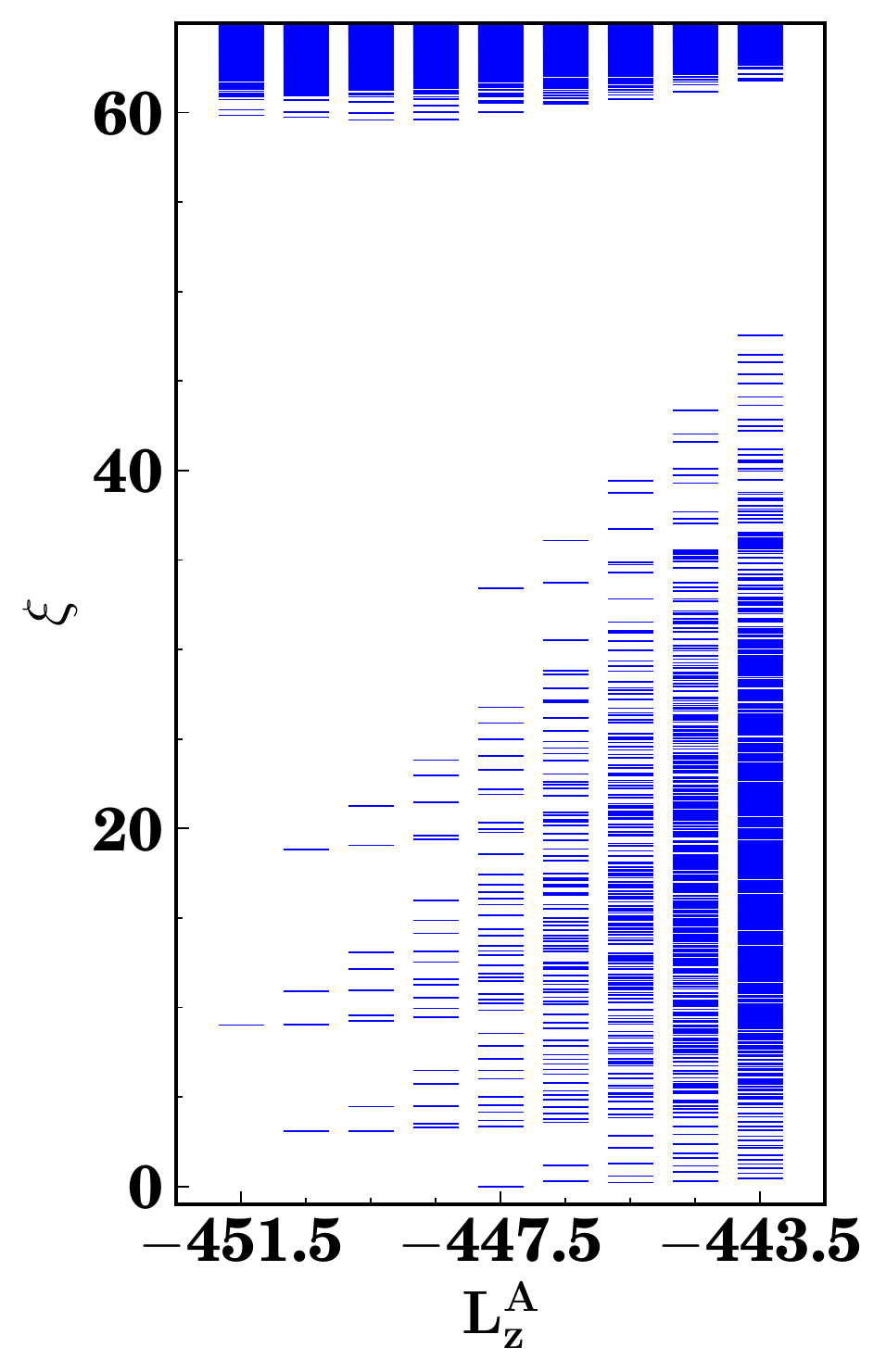}
\caption{{\small Low $L_{z}^{A}$ part of the RSES for the $\nu=2/5$ Jain state, 
 for  $(N,N_A)=(36,18)$ (\emph{left})  and  for $(N,N_A)=(38,19)$ (\emph{right}).
Eigenvalues of $\rho_A$ are calculated up to a single overall scale. This sets the zero of entanglement energy $\xi$. We normalize so that the lowest level shown has $\xi=0$. 
}}
\label{fig:25plots}
\end{center}
\end{figure}
\begin{table}
\vspace*{-7mm}
\begin{tabular}{c}
\begin{tabular}[t]{|c|c|c|c|c|c|c|c|c|c|c|c|c|c|c|}
\hline 
  \backslashbox{$E_{CF}$}{$L_z^A$} & -405 & -404 & -403 & -402 & -401 & -400 & -399 & -398    \\ \hline
   7  &  - & - & - & - & - & - & 1 & 2 \\
   8  &  - & - & 1 & 2 & 5 & 10 & 20 & 36 \\
   9  & 1  & 2 & 5 & 10 & 20 & 36 & 65 & 110 \\
   10 & 1  & 2 & 5 & 10 & 20 & 36 & 65 & 110 \\
   11 &  - & - & 1 & 2 & 5 & 10 & 20 & 36\\
   12 &  - & - & - & - & - & - & 1 & 2 \\ \hline
   Total  & 2 & 4 & 12 & 24 & 50 & 92 & 172 & 296  \\ 
\hline 
\hline 
 \backslashbox{$E_{CF}$}{$L_z^A$} &  \footnotesize{-451.5} &  \footnotesize{-450.5} &  \footnotesize{-449.5} &  \footnotesize{-448.5} &  \footnotesize{-447.5} &  \footnotesize{-446.5} &  \footnotesize{-445.5} &  \footnotesize{-444.5}     \\ \hline
   8  &  - & - & - & - & 1 & 2 & 5 & 10  \\
   9  &  - & 1 & 2 & 5 & 10 & 20 & 36 & 65  \\
   10 & 1  & 2 & 5 & 10 & 20 & 36 & 65 &110  \\
   11 & -  & 1 & 2 & 5 & 10 & 20 & 36 & 65  \\
   12 &  - & - & - & - & 1 & 2 & 5 & 10  \\  \hline
   Total  & 1 & 4 & 9 & 15 & 42 & 80 & 147 & 260  \\ \hline 
\end{tabular}
\end{tabular}
\noindent \caption{Numbers of independent CF-EWF  vs.~$E_{CF}$ for the $\nu=2/5$ Jain state. \emph{Top:} $N=36$ and $N_A=18$, \emph{Bottom:} $N=38$ and $N_A=19$.}
\label{tab:25count}
\end{table}

Fig.~\ref{fig:25plots} shows the RSES of the $\nu=2/5$ Jain state for $N=36$ and $N=38$ particles.  
Observe that the RSES exhibits branches of excitations and that it depends on the $N_A$ parity (similar results have been obtained in studies of edge states in ref.\cite{Jain_edge}). We find that, in the large $N_A$ limit, the RSES counting is given by $2,4,12,24,44,50,92,..$ for even $N_A$  and by $1,4,9,20,42,..$ for odd $N_A$. The counting within each branch is independent of the parity and given by,  $1,2,5,10,20,\ldots$, which corresponds to the counting of a $U(1) \times U(1)$ conformal field theory, or two non-interacting chiral edge bosons. 

This counting can be understood from the point of view of the CF-EWF. The Slater determinants in the EWF which relate to $\chi_2$ can be characterized by the numbers $n_0$ and $n_1=N_{A}-n_0$ of orbitals  filled in the first and second CF LLs, or alternatively in terms of the CF kinetic energy $E_{CF}$. We will take $E_{CF}=\sum_{l} l n_l$, so we work in units of the CF cyclotron energy and set the zero point energy for particles in the lowest CF LL to zero.  
In Table \ref{tab:25count}, we give the number of linearly independent CF-EWF in terms of the CF kinetic energy, $E_{CF}$, for the first $L_z^A$ sectors. 
%
%
Observe that the counting within a row, given by the number of independent CF-EWF with the same kinetic energy, matches the counting of the RSES branches. Moreover, if we interpret every row as a branch we recover exactly the same multi-branch structure observed in Fig.~\ref{fig:25plots}, with the expected counting in the large $N_A$ limit.

Note that because we are considering two CF LL's, the CF-EWF contributing to a single branch, i.e.~the CF-EWF with the same kinetic energy, correspond to states in which the composite fermions are excited only within the same CF LL (inter-LL CF excitations), otherwise we change $E_{CF}$ and we move to another branch. Because inter-LL excitations produce a $U(1)$ counting per LL, in the large $N_A$ limit, this explains the $U(1) \times U(1)$ counting observed in the $\nu=2/5$ RSES branches. 

\begin{figure}[thb]
\begin{center}
\includegraphics[width= 4.2cm, height=8cm]{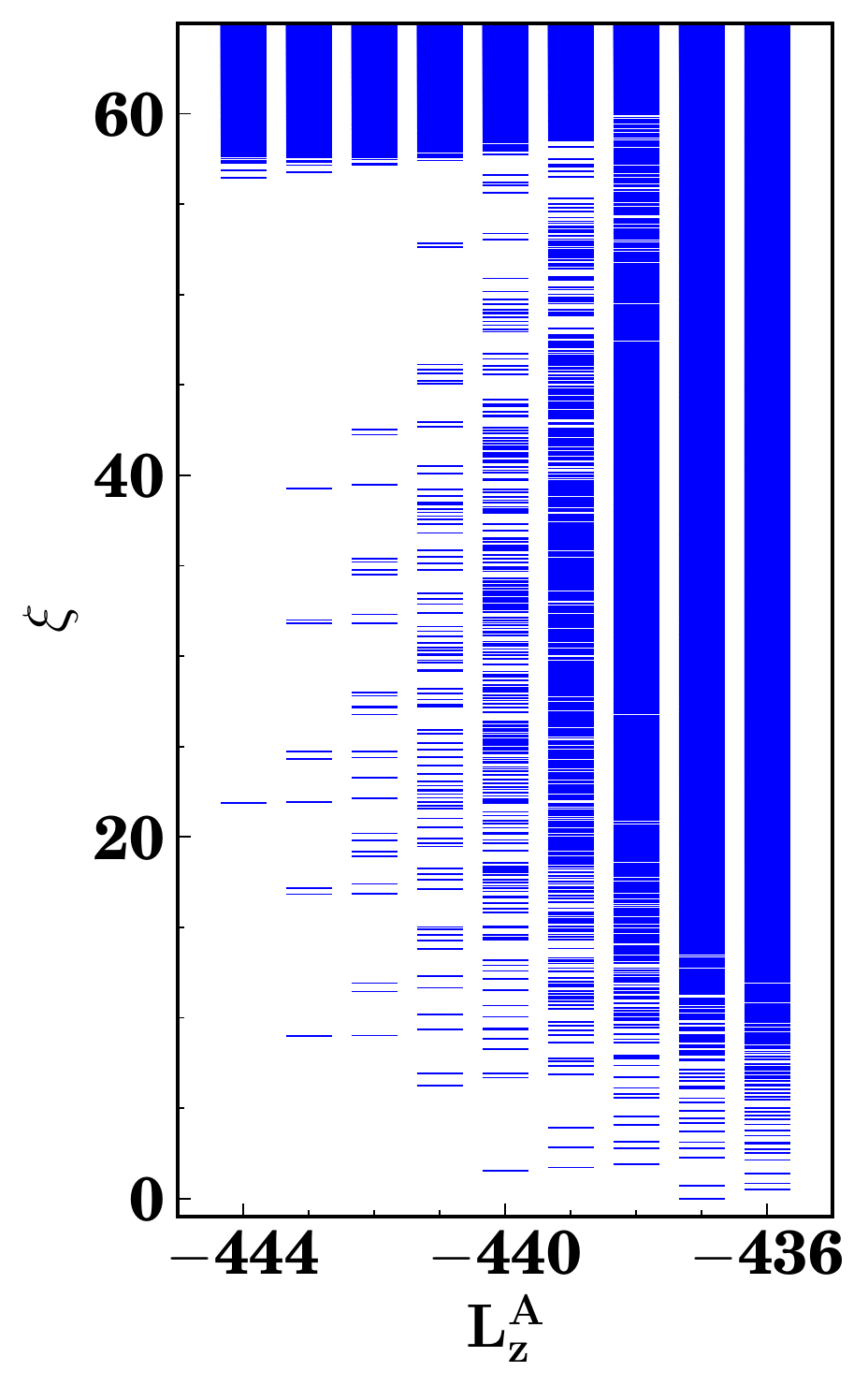}
\includegraphics[width= 4.2cm, height=8.1cm]{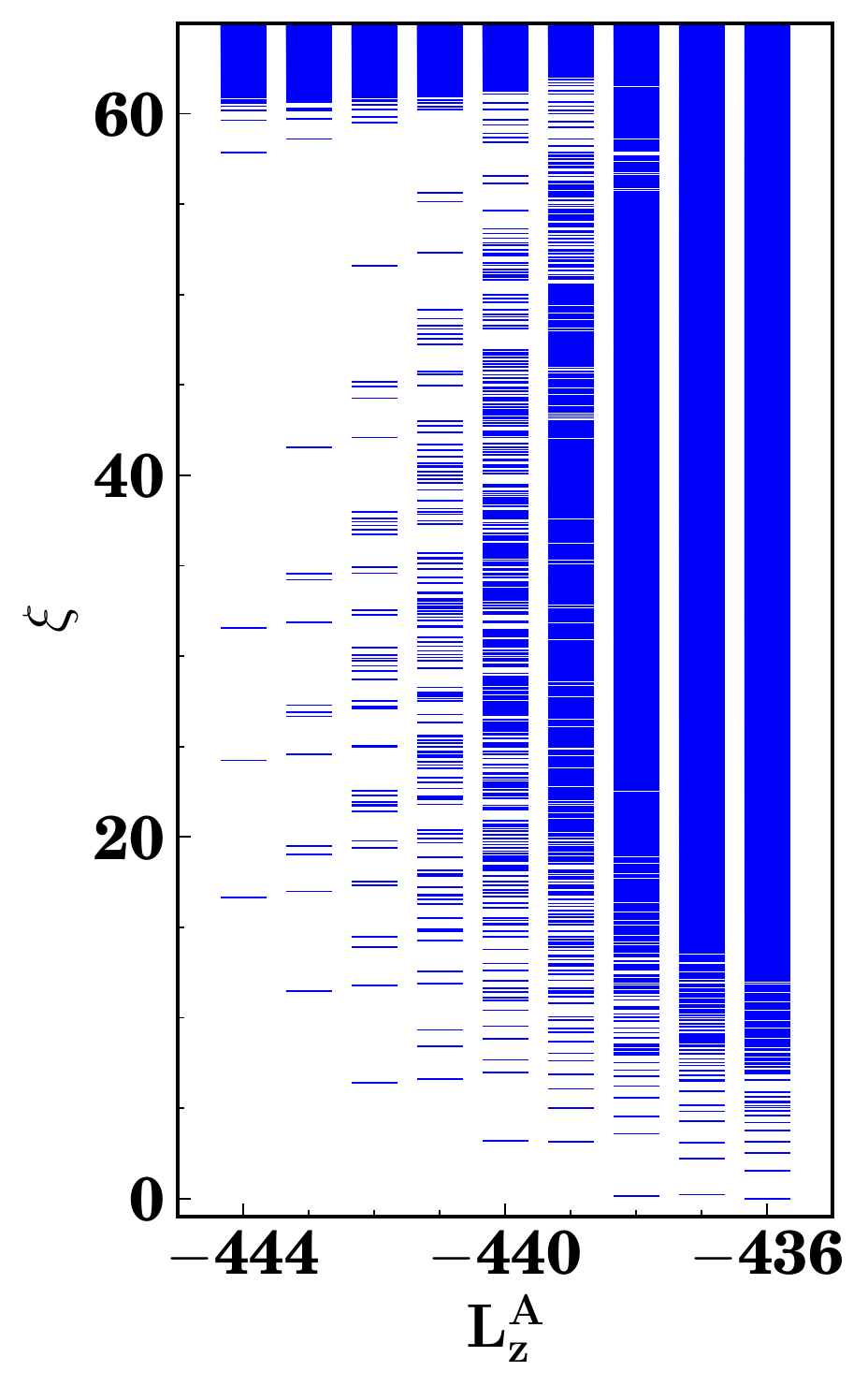}
\caption{{\small RSES counting for the $\nu=3/7$ Jain state for  $(N,N_A)=(42,21)$ (\emph{left})  and for  $(N,N_A)=(39,19)$ (\emph{right}). 
}}
\label{fig:37plots}
\end{center}
\end{figure}
\vspace*{-8mm}
\begin{table}[htb]
\centering
\resizebox{9.0cm}{!}{
\subfloat{
\begin{tabular}[t]{|c|c|c|c|c|c|c|c|c|c|c|c|c|c|c|}
\hline 
\backslashbox{$E_{CF}$}{$L_z^A$}  & -515.5 & -514.5 & -513.5 & -512.5 & -511.5      \\ \hline
   19 &  - & - & - & - &  1  \\
   20  &  - & - & - & 2 & 6  \\
   21 & -  & 1 & 3 & 9  & 24\\
   22 & -  & 2 & 6 & 18 & 44 \\
   23 &  1 & 3 & 9 & 24 & 57  \\
   24 &  - & 2 & 6 & 18 & 44  \\ 
   25 &  - & 1 & 3 & 9  & 24 \\
   26 &  - & - & - & 2  & 6 \\
   27 &  - & - & - & -  & 1 \\ \hline
   Total  & 1 & 9 & 27 & 82 &207   \\ 
\hline 
\end{tabular}
}
\subfloat{
\begin{tabular}[t]{|c|c|c|c|c|c|c|c|c|c|c|c|c|c|c|}
\hline 
   \backslashbox{$E_{CF}$}{$L_z^A$} & -444 & -443 & -442 & -441  & -440    \\ \hline
   17 & -  & - & - & -  & 1 \\
   18 & -  & - & 1 & 3  & 10\\
   19 & -  & 1 & 4 & 12 &  31\\
   20 &  1 & 3 & 10 & 25 &  61  \\
   21 &  1 & 4 & 12 & 31 & 73  \\ 
   22 &  1 & 3 & 10 & 25 &  61 \\
   23 &  - & 1 & 4 & 12  &  31\\
   24 &  - & - & 1 & 3  &  10\\ 
   25 & -  & - & - & -  & 1 \\ \hline
   Total  & 3 & 12 & 42 & 111 & 279  \\ 
\hline 
\end{tabular}
}
}
\caption{number of independent  CF-EWF vs. $E_{CF}$ for the $\nu=3/7$ Jain state. \emph{Left:}  $N=42$ and $N_A=21$. \emph{Right:} $N=39$ and $N_A=19$}
\label{tab:37count}
\end{table}

For the $\nu=3/7$ Jain state we also find a RSES with many branches of excitations (see Fig.~\ref{fig:37plots}).  As shown in Table~\ref{tab:37count}, for $N_A \mbox{ mod } 3=0$, the spectrum presents two types of branches with countings $1,3,9,24,..$ and $2,6,18,..$. For any other value of $N_A$, there are two type of branches, with countings $1,3,10,25,..$ and $1,4,12,31,..$. Again, these countings are as predicted by the CF kinetic energy of the CF-EWF. 
Note that now we are considering three CF LL in our construction and as a result, we have CF-EWF with the same $E_{CF}$ but with different numbers of CF's in the CF LL's (e.g. for $N_{A}=3$, we may have $1$ CF in each CF LL or $3$ CFs in the middle CF LL) Therefore the counting of the branches for large $N_A$ exceeds the counting that would be expected from $3$ independent chiral boson edge modes.

Explicit generating functions for the numbers of independent CF states with given $E_{CF}$ and $L_z$ will be presented in a forthcoming work~\cite{inprep}. There we will also present results on the RSES of Jain states with reverse flux attachment. The EWF and CF-EWF can also be applied to these states, but the resulting countings are more difficult to obtain, since the LLL projection $\mathbf{P}$ now has a nontrivial effect, even as $N_{A}\rightarrow\infty$.

\noindent \textbf{Acknowledgments:} JKS and IDR were supported in part by Science Foundation Ireland Principal Investigator award 08/IN.1/I1961.  IDR acknowledges SISSA and INFN for financial support. SCD and SHS were supported by EPSRC Grants No. EP/I032487/1 and No. EP/I031014/1. We acknowledge use of the Hydra cluster computer at the Rudolf Peierls Centre for Theoretical Physics and we thank A.Cappelli for helpful discussions and Jonathan Patterson for technical assistance with running the numerical calculations.


\end{document}